# One-Step Hydrothermal Synthesis of Sb$_2$WO$_6$ Nanoparticle towards Excellent LED Light Driven Photocatalytic Dye Degradation

Devdas Karmakar[a], Sujoy Kumar Mandal[a], Sumana Paul[b], Saptarshi Pal[c,d], Manik Pradhan[c,e], Sujoy Datta[a], Debnarayan Jana[a]*

a. Department of physics, University of Calcutta, 92 A.P.C. Road Kolkata-700009, India
b. Department of Physics, Indian Institute of Technology, Guwahati-781039, India
c. Technical Research Centre, S. N. Bose National Centre for Basic Sciences, JD Block Sector-III, Salt Lake City, Kolkata - 700106, India
d. Department of Physics, Institute of Applied Sciences and Humanities, GLA University, Mathura 281406, Uttar Pradesh, India
e. Department of Chemical, Biological and Macromolecular Sciences, S. N. Bose National Centre for Basic Sciences, Salt Lake, JD Block, Sector III, Kolkata 700106, India

Corresponding author Email id: djphy@caluniv.ac.in

**Abstract:**

Pristine Antimony tungstate nanoparticles prepared via a simple hydrothermal process showcase interesting photocatalytic efficiency, degrading Methylene Blue (MB) completely in 180 min under visible light. In this study, the impact on crystalline quality and related optical properties as well as photocatalytic efficiency of antimony tungstate due to temperature variation during hydrothermal synthesis are explored. While X-ray diffraction (XRD) shows the polycrystalline nature of all synthesized samples, however a systematic increase in crystallite size is revealed by analysing the XRD peak broadening. XRD spectra are further examined by Rietveld analysis showing a change in unit cell volume. Additionally, the overall changes in the corresponding grain size and micro-strain developed in the crystals are determined using the Williamson-Hall plot. Moreover, significant variations in few Raman modes are observed with increasing synthesis temperature. A notable modification in the optical band gap as determined from the absorbance of the UV-Vis spectra is perceived with the change in synthesis temperature within the range of ~2.38-2.57 eV. Further, the photoluminescence measurement indicates that the synthesized antimony tungstate is weak luminescent material with a band-to-band emission at ~468 nm. Finally, photocatalytic efficiencies of the samples are ascertained to change with the synthesis parameter, estimated by decomposing methylene blue (MB), highest degradation rate constant (k) value is observed as 0.015 min$^{-1}$ for the sample prepared at 180 °C. While the sample with the highest efficiency is also applied for degrading the Rhodamine B (RhB) and Potassium Dichromate ($K_2Cr_2O_7$) under visible light irradiation.

## 1. Introduction:

According to the World Bank report, textile and dyeing industries contribute about 17–20% of industrial water pollution [1]. These discharges contain non-biodegradable, highly toxic colour pigments that are harmful to living organisms [2,3]. Hence, several strategies are adopted to invalidate this serious problem, including ozonation [4], bio adsorption [5], membrane filtration [6], removal by ion exchange [7], adsorption [8], photocatalytic degradation [9,10], catalytic reduction [11], aerobic or biological treatment [12] and coagulation [13]. Other than carcinogenic dyes, heavy metals are also hazardous to human health. For example, Chromium exists mainly in hexavalent Cr(VI) and trivalent Cr(III) forms. Among these, Cr(VI) is a type I carcinogenic causing chronic ulcers, gastrointestinal tract damage while Cr(III) is one of the essential micronutrient for human health [14,15,16]. According to World Health Organization (WHO), the maximum allowable limit for the Cr(VI) in drinking water is 0.05 ppm [14, 17]. Therefore, the reduction of Cr(VI) to the less harmful Cr(III) is very demanding.

In order to nullify the lethal effect of these carcinogenic dyes photocatalysis is considered as a favourable technology for industrial wastewater treatment techniques owing to its environmentally friendly method, low cost and most importantly lack of secondary pollution [18,19]. Among the most studied materials $TiO_2$, ZnO, ZnS, CdS and other metal oxides hold the topmost position [9, 20, 21, 22-27]. Besides using different forms of pristine materials, several strategies are being used by the scientific community towards further developing this field by doping, heterojunction formation and defects formation.

Along with the above-mentioned semiconductors, $Sb_2WO_6$ is also gaining more and more attention due to its instinct potential. Being a crucial member of Aurivillius family with a typical layer structure, consists of perovskite-like layers of $[WO_4]^{2-}$ sandwiched between two $[Sb_2O_2]^{2+}$ layers [28-33]. While the perovskite-like layers of $[WO_4]^{2-}$ in $Sb_2WO_6$ possess significantly distorted structures than other members of Aurivillius family. These unsaturated sites in the structure give rise to an internal electric field acting positively for charge carrier separation and this may be one of the reasons promoting its use as photocatalyst [34]. Moreover, these unsaturated sites may be favourable for the adsorption of relevant molecules and enhance significantly the interfacial interactions for efficient exchange of active particles.

Besides $Sb_2WO_6$ has attracted strong interest in the photocatalytic field because of its high visible light absorption performance and narrow band gap, lying in the visible region [35-38]. Thus, this material has the internal potential to be used as photocatalyst. Several authors have applied different strategies to improve the efficiency of this sample. Ren *et. al.* prepared $Sb_2WO_6$/g-$C_3N_4$ composite for oxidation of NO [35]. While Rafiq *et. al.* processed rGO-$Sb_2WO_6$ composite for degrading RhB under visible light [36]. Chen *et. al.* prepared Ag loaded $Sb_2WO_6$ for degrading different azo dyes under ultraviolet (UV) and visible light [37]. Pt doped $Sb_2WO_6$ has been synthesized by Shi *et. al.* for reducing nitrobenzene to aniline [28] . But pristine $Sb_2WO_6$ has never shown a significant photocatalytic effect due to rapid electron hole pair recombination [38,39]. While our previous study using Density Functional Theory (DFT) indicates that $Sb_2WO_6$ has appreciable absorption in the visible region [40]. Though Bi et. al. have prepared pristine $Sb_2WO_6$ in hydrothermal process at different interval of time [41], however the change in the photocatalytic efficiency under different hydrothermal temperature, has not been studied still now. Thus with an aim to optimize the reaction temperature, a simplified method is used here for preparing photocatalysts using Antimony tungstate nanoparticles (ATN).

Herein, ATN are obtained by one step facile hydrothermal process with temperature optimization as the most important parameter. Interestingly, it is observed that temperature can significantly tune the properties of ATN promoting a particular sample with significant photocatalytic properties. The main aim of this study is that by investigating optimal synthesis conditions to elucidate the most significant active species for the highest photo decolorization reactions. Primarily, the most efficient sample is selected by degrading MB and finally it is also applied for degrading RhB and Potassium dichromate. The results of our study will shed light on further preparation of a more effective and economical synthesis strategy for photocatalyst based on ATN.

## 2. Experimental Section
### 2.1. Materials

The ingredient Antimony Chloride ($SbCl_3$) is purchased from Merck Specialities private limited with 99.0% purity. Sodium Tungstate ($Na_2WO_4, 2H_2O$) is taken from Spectrochem Pvt. Ltd. with a purity of 99.0%. All the chemicals are of analytical grade and used without further purification. Doubled distilled water was used for synthesis purposes.

**2.2. Synthesis Procedure**

Antimony Tungstate is prepared using antimony chloride and sodium tungstate as precursor via a typical hydrothermal process. At first, 0.8 m mole $SbCl_3$ and 0.4 m mole $Na_2WO_4, 2H_2O$ are separately dissolved in two 100 ml beakers each containing 22 ml water at pH 6.7. To prepare a well-dispersed solution the beakers were separately sonicated for 15 minutes and further stirred vigorously at 400 rpm for 30 minutes. After that, the $Na_2WO_4, 2H_2O$ solution is poured into the $SbCl_3$ solution drop by drop while stirring. It is further stirred for another one hour to make a homogeneous solution. The mixture was then transferred to a Teflon-lined autoclave of 55 ml volume capacity and placed in a preheated oven for 12 hours. A series of samples were prepared at different hydrothermal temperatures of 120 $^oC$, 140 $^oC$, 160 $^oC$, 180 $^oC$ and 200 $^oC$. After 12 hours of heating, the samples were allowed to cool naturally. The samples were then collected via centrifugation and washed repeatedly with the ethanol-water mixture to remove the unreacted parts, then finally dried overnight at 70 $^oC$. Hereafter, the samples are denoted by S120, S140, S160, S180 and S200 respectively.

**2.3. Photocatalytic Dye Degradation Activity**

The photocatalytic experiments are carried out by degrading methylene blue in an aqueous solution under 30 watt visible Light Emitting Diode (LED). In a typical experiment [9], 5 mg sample is added with double distilled water which is sonicated for 15 minutes after that some dye is added from stock the solution totalling volume of 25 ml with a dye concentration of $10^{-5}$ Mole. After adding the dye, the solution is kept in dark for 30 minutes to attain dye adsorption –desorption equilibrium. Then the solution is placed under a visible LED keeping the initial distance of the upper surface of the solution from the light at 10 cm. To track the remaining dye amount 2ml aliquots are collected at the 30 minute interval and centrifuged at 8000 rpm for 4 minutes to separate the sample. Finally, dye concentration in the extraction is monitored by taking UV-Vis adsorption. The optimized sample is also used for degrading Rhodamine B (RhB) and potassium dichromate.

**2.4 Characterization**

The X-ray diffraction pattern is obtained from a Bruker AXS D8 Advanced equipment with Cu-K$_\alpha$ radiation (λ=1.5406 Å). The XRD scan is taken in the 2θ range 15º-70º at room temperature with a scan rate of 0.03 deg/sec. The UV-Vis spectra were recorded by the Shimadzu UV-1800 UV-visible Scanning Spectrophotometer. The room temperature Raman spectra were registered from Jobin Yvon Horiba LabRAM HR Evolution spectrometer using 532 nm laser. To study the surface morphology of the samples, Scanning Electron Microscope (SEM) images have been recorded by using ZEISS EVO 18 instrument while the elemental composition is determined from ZEISS SmartEDX. Further, Transmission electron microscope (TEM) images are recorded by high-resolution transmission electron microscope (HRTEM) [UHR-FEG TEM, JEM-2100F, Jeol, Japan]. The photoluminescence spectra were measured by FluroMAX3 instrument with a xenon lamp as the excitation source to use different excitation lines.

## 3 Results and Discussion

### 3.1 Structural and Morphological Analyses

#### 3.1.1 XRD Pattern Analyses

The structural information of any material is obtained from the X-Ray Diffraction (XRD) pattern. **Figure 1** shows the XRD pattern of the synthesized $Sb_2WO_6$ (JCPDS no. 47.1680) [18,39]. The peaks as shown in **Figure 1** with 2*θ* at 20.14, 27.07, 29.14, 32.94, 36.55, 40.32, 47.64, 49.89, 53.34 and 55.51 correspond to the crystal planes with (hkl) values (011), (012), (003), (20-1), (020), (202), (023), (22-1), (310) and (222) respectively [42]. These peaks are related to the pure orthorhombic phase of $Sb_2WO_6$ [43]. There are no other impurity peaks present in the samples. Therefore, even at the lowest temperature of this series i.e., 120 ºC, the pure phase of antimony tungstate is noted. However, with the increase in temperature, the peaks become sharper, indicating the crystals are getting more crystallinity. Interestingly, the intensity of the (011) plane is found to be gradually increased up to S180 but almost gets disappeared for S200. This observation manifests that the (011) plane is no more a preferred plane at this synthesis condition (synthesis temperature: 200 ºC).

The grain size i.e., the coherent scattering region (CSR) in a direction to perpendicular to the diffraction plane is inversely proportional to the Full Width Half Maxima (FWHM) of the XRD peaks. The strain development inside the crystal structure is due to the presence of various types of defects caused by the synthesis environment. Different types of defects, mainly plane dislocation ones arise due to strain developed giving rise to the broadening of the diffraction peaks. Thus, the broadening of the XRD peaks is contributed by the finite size as well as micro-strain developed inside the crystals. The individual contribution of these two effects however can be calculated by using Williamson-Hall method [44, 45]. According to this, the total FWHM of XRD peaks is equal to the sum of the FWHM of the above two factors taken individually

$$\beta_{total} = \beta_{size} + \beta_{strain} \quad \text{.........................} \quad (i)$$

Here, the Uniform Deformation Model (UDM) is considered according to which the strain developed due to the crystal deformation is uniform throughout the crystallographic direction and isotropic in nature [44,46].

The broadening due to crystal size is given by [35], $\beta_{size} = \frac{k\lambda}{D\cos\theta}$ ....... (ii)

where, 'λ' is the wavelength of X-rays, 'D' represents average grain size, and 'k' is the shape factor. In the similar way, the expression involving the micro-strain (ε) term [44, 45] will be,

$\beta_{strain} = 4\varepsilon \tan\theta$ ......... (iii)

Thus, the micro-strain (ε) can be calculated from the following equation:

$$\beta_{total} = \frac{k\lambda}{D\cos\theta} + 4\varepsilon \tan\theta \quad \text{....................} \quad (iv)$$

$$\beta_{total} \cos\theta = \frac{k\lambda}{D} + 4\varepsilon \sin\theta \quad \text{................} \quad (v)$$

**Figure 1(b)** represents the W-H plot of all the samples. The negative slope in the W-H plot indicates the presence of compressive strain in the crystal [47]. All the calculated strains are arranged in **Table I**. These types of changes in the unit cell parameters are also observed in some earlier studies [48].

**Table I:** List of average crystallite size and strain induced in the sample as calculated from the XRD pattern using the W-H plot.

| Samples | Average crystallite size (nm) | Strain ($\varepsilon$) |
|---|---|---|
| S120 | 33.8 ± 8.5 | -0.001376 ± 0.000977 |
| S140 | 38.4 ± 10.3 | -0.001042 ± 0.000984 |
| S160 | 41.7 ± 09.9 | -0.001269 ± 0.000745 |
| S180 | 44.0 ± 12.0 | -0.001187 ± 0.000848 |

|  | S200 | 46.0 ± 11.7 | -0.001149 ± 0.000734 |
|---|---|---|---|

Moreover, Rietveld analysis is carried out to know about the crystal cell parameters. The profile matching of Rietveld analysis is done with the space group symmetry P1 of the triclinic lattice. The values of the cell parameters are close consistent with the values given by Castro *et al.* [49]. It is interesting to note that with the increase in the synthesis temperature, the cell volume primarily increases up to S160, but a decrease of the same is observed afterwards similar to the results obtained by Xiuqin *et al.* [50]. The reason for the increase in crystal size may be attributed to the increase in synthesis temperature.

Since Rietveld analysis is done by the fitting of an experimental XRD pattern, it is important to have a numerical figure of merit to quantify the quality of the fit between the calculated and experimental pattern. Here, the parameters $\chi^2$ is considered to represent the goodness of fit are 1.37, 1.55, 1.68, 1.29 and 1.57 for the samples S120 to S200 respectively. The results indicate that our Rietveld analysis is quite well fitted with the experimental data, which can be observed in **Figure** 1(c). The cell parameters as well as cell volume obtained from the Rietveld analysis are tabulated in **Table II**.

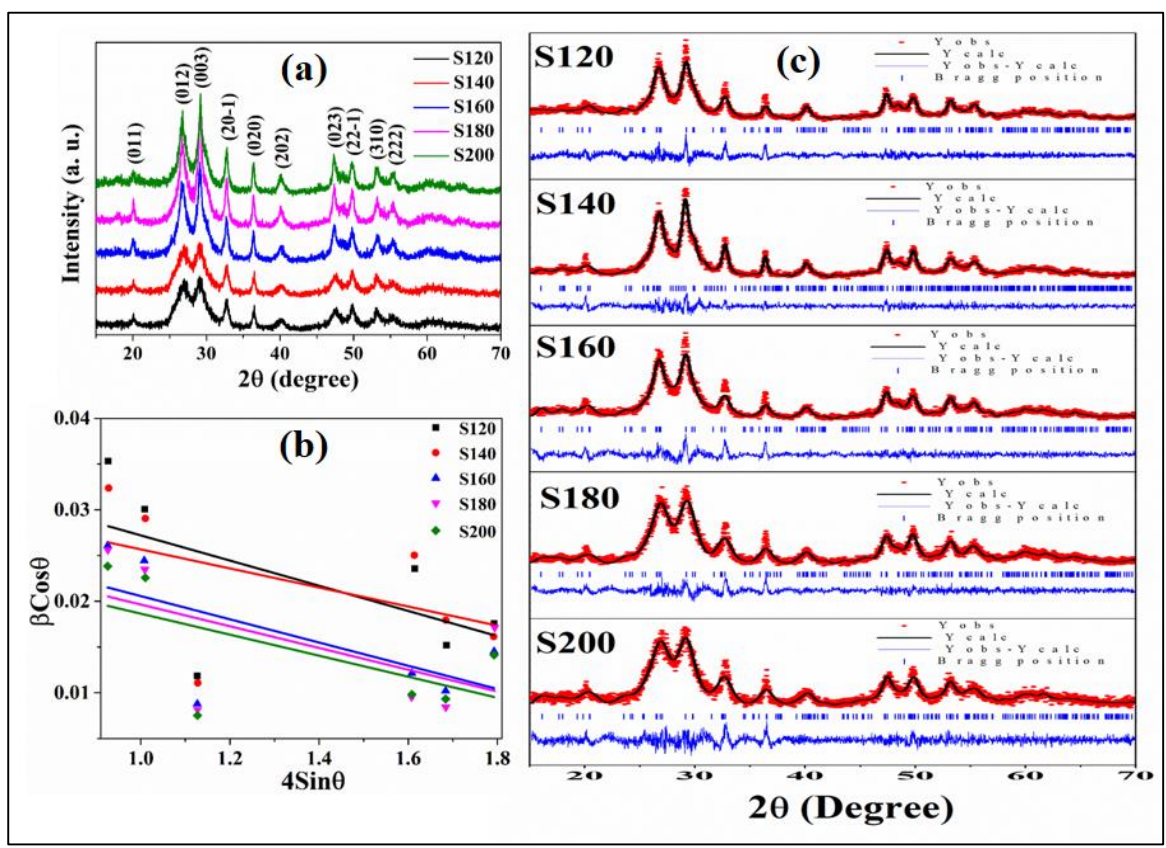

**Figure 1: (a)** XRD pattern, (b) Williamson-Hall plot and (c) Rietveld analysis of the samples S120, S140, S160, S180 and S200.

**Table II:** Cell parameters along with unit cell volume obtained from the Rietveld analysis of the samples S120, S140, S160, S180, and S200.

| Sample | a (Å) | b (Å) | c (Å) | α | β | γ | Volume (Å$^3$) |
|---|---|---|---|---|---|---|---|
| S120 | 5.5432 | 4.9211 | 9.2601 | 90.132 | 97.45 | 90.312 | 250.4637 |
| S140 | 5.5614 | 4.9273 | 9.2234 | 89.873 | 96.889 | 90.536 | 250.9144 |
| S160 | 5.5579 | 4.9288 | 9.2447 | 90.246 | 97.012 | 89.634 | 251.3450 |
| S180 | 5.5576 | 4.9266 | 9.2325 | 89.922 | 96.855 | 90.441 | 250.9709 |
| S200 | 5.5548 | 4.9262 | 9.2366 | 90.007 | 97.051 | 90.369 | 250.8368 |

Now, from the W-H plot, a tendency of decrease in compressive strain is observed. This change in compressive strain in crystal might arise due to the change in shape, size, and morphology with the variation of synthesis conditions. Moreover, from Rietveld analysis a non-monotonic cell volume variation is found with increasing temperature. The maximum cell volume is observed in the case of sample S160. Thus, initially there is an increase in the cell volume with the decrease in the magnitude of the compressive strain up to sample S160.

### 3.1.2 SEM Images Analyses

The morphology of the ATN are studied by SEM images. **Figure 2(a-c)** depicts the SEM images of three samples synthesized at relatively higher temperatures, S160, S180 and S200 respectively. In **Figure 2(a)**, the SEM image of S160 sample is shown and the observed average grain size of the material is around 50 nm. The particles are almost uniform and they are of spherical shape. **Figure 2(b), (c)** represents the SEM images of samples S180 and S200 respectively. No notable changes in the grain size and shape can be seen for these two samples when compared with S160. In general, an increase in crystallite size as revealed from XRD is directly correlated to the grain growth. However, here, the variation of crystallite size is not in higher extent thus it should not reflect in the SEM images as grain growth. Thus, the outcomes of XRD and SEM are consistent. Moreover the elemental composition of the sample S180 is shown in Figure 2(d) where the insert image shows the selected area over which EDAX is done. The constituting elements and their corresponding percentage are given in the **Table III**.

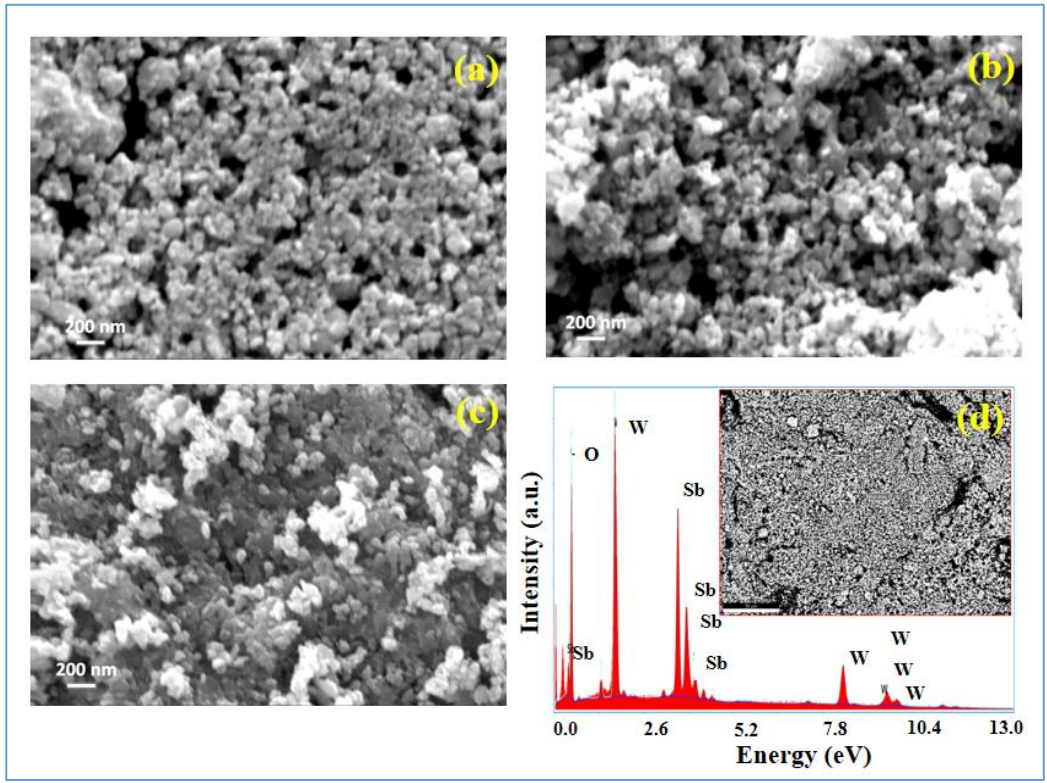

**Figure 2.** SEM images of the samples for (a) S160, (b) S180 and (c) S200 samples (d) EDX spectrum of S180.

Table III: Amount of the constituting elements present in the sample S180.

| Element | Weight% | Atomic% | Error% |
| --- | --- | --- | --- |
| O (K) | 16.9 | 64.7 | 8.4 |
| Sb (L) | 44.8 | 22.5 | 3.3 |
| W (L) | 38.3 | 12.2 | 7.4 |

### 3.1.3 TEM analysis:

To get a better understanding of the morphology of the nanoparticles along with a more accurate size of the sample S180, transmission electron microscopy analysis is carried out. **Figure** 3 (a) shows the TEM images of the mentioned sample and a magnified view of the same is shown in **Figure 3(b)**. From the staking of more than one particles, as shown in **Figure 3(b),** the particle size measured from this image lies near 50 nm. The HRTEM image of the sample is shown in Figure 3 (c). The FFT pattern, as obtained from the yellow square region of **Figure 3(c)** shows the ($\bar{3}$11) plane of the sample. The reconstructed HRTEM image by masking the yellow circles in **Figure** 3(d) is shown in **Figure** 3(e) and the corresponding *d* spacing is 0.34 nm. In the SAED pattern, a polycrystalline nature of the sample is observed which is shown in Figure 3(f). Different planes corresponding to the different diffraction planes are shown with yellow circles.

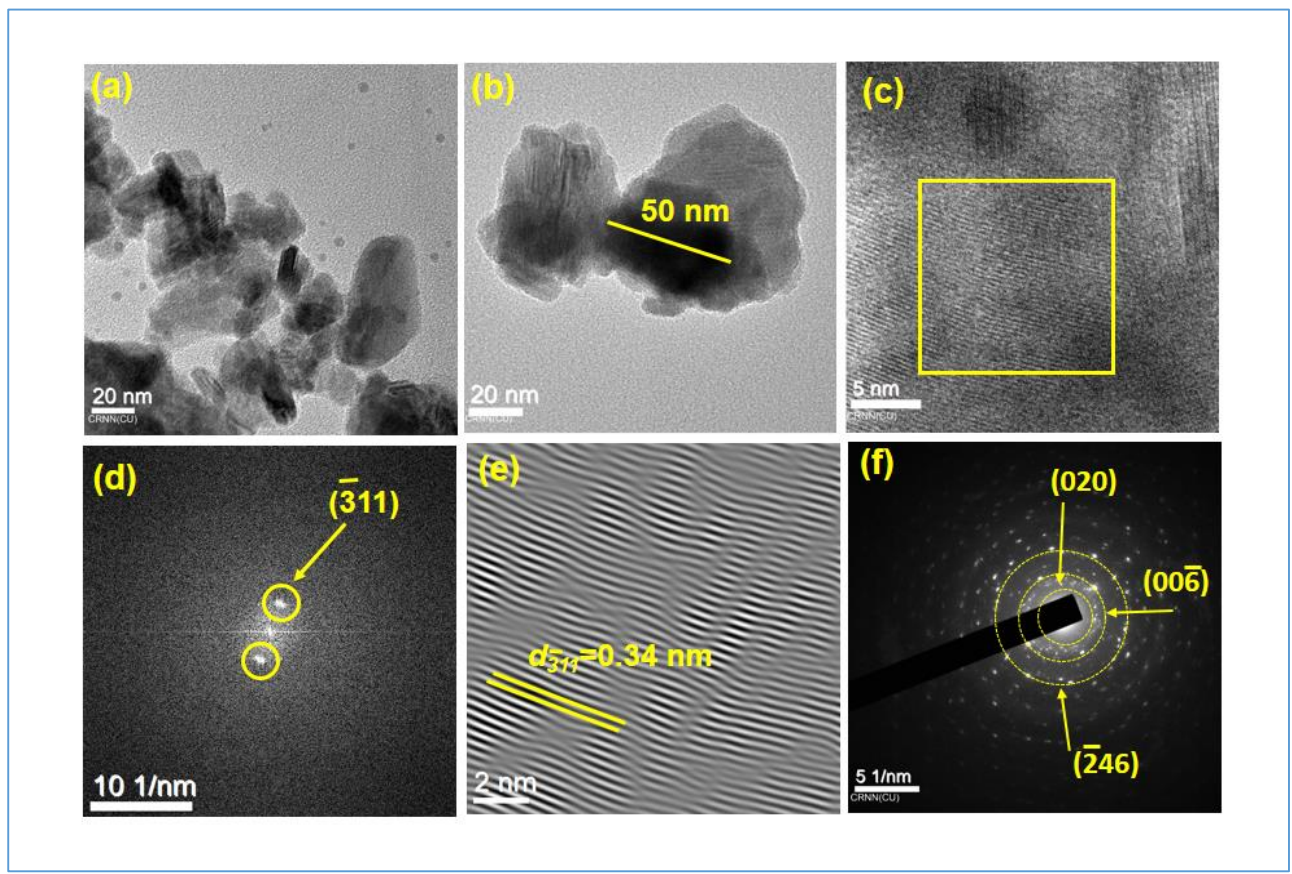

**Figure** 3: (a) and (b) TEM images at different sites, (c) HRTEM, (d) FFT image of the selected region in (c), (e) Closer look to the HRTEM image and (f) SAED pattern of S180 sample.

### 3.1.4 *Raman Spectra Analyses*

The information about molecular and crystalline vibrational modes can be obtained from the Raman spectra. **Figure 4(a)** represents the Raman spectra of all the synthesized samples under discussion. The report on $Sb_2WO_6$ Raman spectra is scarce in the literature. Moreover, previously reported Raman spectra are not well resolved in most of the cases, even a careful analysis to assign the Raman modes has not been attempted [38,51]. In this context, Raman spectra analysis of other crystals with similar Aurivillius structures could be helpful, as $Sb_2WO_6$ and $Bi_2WO_6$ both belonging to the Aurivillius framework with similar structure, sharing the layers of $WO_6$ octahedra as common. However, the presence of $[Sb_2O_2]$ in spite of $[Bi_2O_2]$ should give a different characteristic signature. So, it is expected that the Raman peaks corresponding to the $WO_6$ should possess the same properties and that of the $[Sb_2O_2]$ layer may showcase some difference. The simplest types of Aurivillius structures have six Raman active modes ($2A_{1g}+B_{1g}+3E_g$) with nine infrared (IR) active vibrations modes [52, 53]. Out of these, $A_{1g}$ is the symmetric vibration of the $WO_6$ octahedra. Further, stretching and bending vibration of ($Bi_2O_2$) layers give ($B_{1g}+E_g$) mode and translation of $Bi^{3+}$ ion is related to ($A_{1g}+E_g$) mode [54]. The major peaks below 200 $cm^{-1}$ may be due to the translation of Sb and W atoms [55]. Thus, the two peaks at 80 $cm^{-1}$ and 164 $cm^{-1}$ could be due to the translational motion of the metal ions [56]. Moreover, the crystals $Bi_2WO_6$, $Sb_2MoO_6$ and $Bi_2MoO_6$ belong to the Aurivillius framework with the n=1, where n is the number of perovskite layers . At the same time, these materials also show close crystal symmetry with the $Sb_2WO_6$. Thus, a close relation of their Raman spectra can be expected [53] with small changes due to their variation in the mass of the constituting atoms and their radii [57]. The peak for antisymmetric bridging modes of W (Mo) atoms for $Bi_2WO_6$ ($Bi_2MoO_6$) arises at 714 $cm^{-1}$ (713 $cm^{-1}$) [54, 58], while a prominent peak for $Sb_2MoO_6$ is found to be reported at 720 $cm^{-1}$ [59]. Comparing these peaks, the Raman modes at 715 $cm^{-1}$ for $Sb_2WO_6$ could be linked with the anti-symmetric bridging modes of W atoms. Again, for $Bi_2WO_6$ there are two nearby peaks at 789 $cm^{-1}$ and 820 $cm^{-1}$ which are ascribed to the antisymmetric and symmetric $A_g$ modes of O-W-O bonds. Similarly, $A_g$ modes for asymmetric stretching of $MoO_6$ octahedra of $Bi_2MoO_6$ are obtained at 793 $cm^{-1}$ and 840 $cm^{-1}$. Although a broad peak at 810 $cm^{-1}$ can be seen for $Sb_2MoO_6$, however the nature and reason for the broadness of this specific Raman mode is not further studied [59]. Here, the broad peak around 880 $cm^{-1}$ is deconvoluted with the Lorentzian curve and the result is shown in **Figure 4(b,c)**. The deconvoluted images indicate the presence of two peaks at 846 $cm^{-1}$ and 884 $cm^{-1}$. These peaks can be assigned due to the vibrational modes of antisymmetric and symmetric stretch of the apical oxygen atoms of the $WO_6$ octahedron. The ratio of the intensity of the peaks at 846 $cm^{-1}$ relative to the peak at 884 $cm^{-1}$ for all the samples are given in **Figure 4(d)**. Interestingly, an almost linear increase with respect to the temperature of synthesis of the said ratio can be seen. Therefore, Raman analysis reveals significant changes in the internal crystal structure with the change in synthesis temperature.

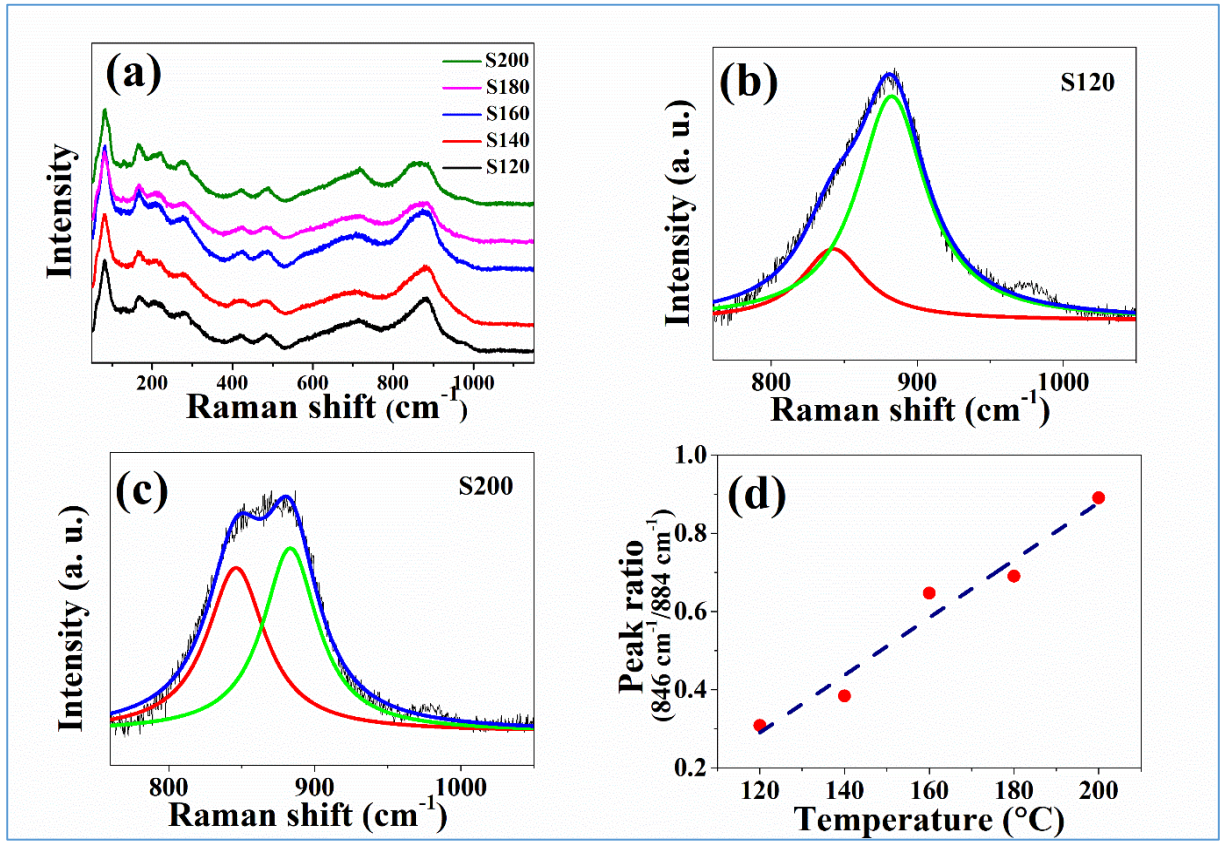

**Figure 4.** **(a)** Raman spectra of the as-prepared samples from S120 to S200. Deconvoluted Raman peak at 880 cm$^{-1}$ into two Lorentzian curves for the samples (b) S120, (c) S200. (d) The peak intensity ratio (846 cm$^{-1}$/884 cm$^{-1}$) of the two deconvoluted peaks.

### 3.2. Optical property of the Synthesized Samples

#### 3.2.1 *UV-Visible Absorbance Spectra Analyses*

**Figure 5(a)** shows the UV-visible absorbance spectra of all ATN. The measurements are carried out by dispersing the powder samples in aqueous medium with same concentration. A long band-tailing for all the samples is observed, however the tailing is not same for all the samples. The possible reason for this tailing could be the presence of different surface defect states which can create sub band gap levels. Moreover, the interaction between such surface defects state and solvent water can trigger this tailing. As no sharp band to band absorption edge is present for all the samples, here, differential plot method (dA/dλ with respect to λ) is used to evaluate the optical band gap of the system. The point at which the absorbance intensity just starts increasing sharply [see **Figure 5(a)**], i.e., the point of inflexion [see **Figure 5(b)**] will give the effective band edge transition. The associated bandgap is calculated by the formula $E_g=1240/\lambda$ eV where λ is the band-edge transition wavelength in nm. The bandgap determined from this plot for the samples S120, S140, S160, S180 and S200 are respectively 2.57 eV, 2.45 eV, 2.39 eV, 2.38 eV and 2.41 eV. This results are consistent with our previous study [40]. It is interesting to note that there is a significant change in the bandgap due to the variation of the synthesis temperature. Bandgap decreases from ~2.57 eV for S120 to the lowest value of ~2.38 eV for the sample S180. Further, the bandgap increases again for S200 and takes an intermediate value. Thus, the temperature plays a significant role in tuning the effective bandgap of the samples. This bandgap tuning may be attributed to the cationic disorder and large distortion of the $WO_6$ octahedra [60].

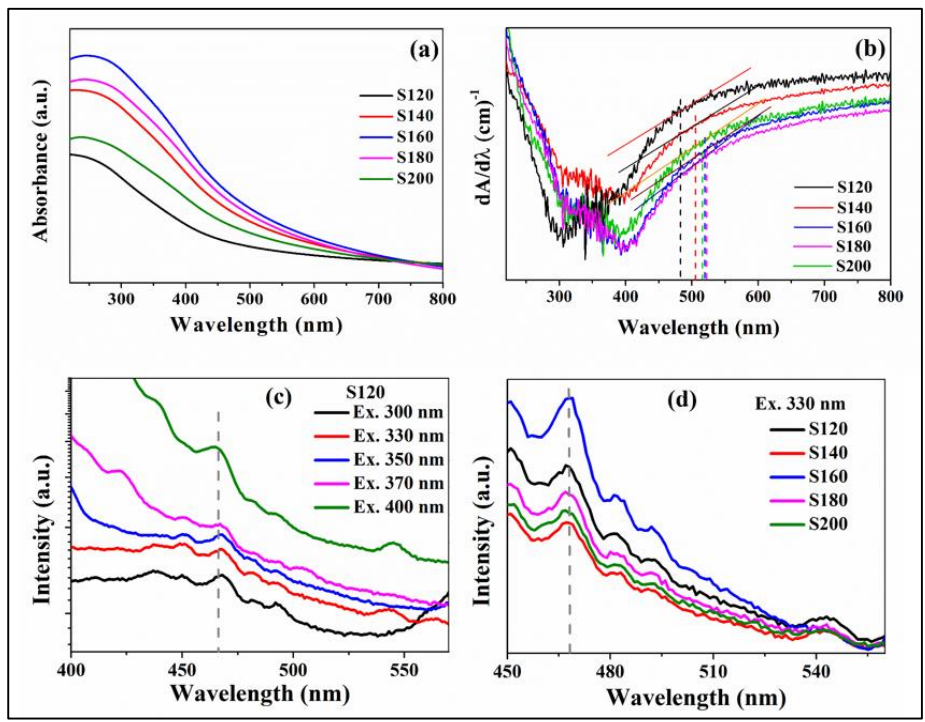

**Figure 5:** (a) UV-visible absorbance spectra of all $Sb_2WO_6$ samples and (b) first derivative of the absorbance spectra with respect to the wavelength is shown. The point of inflexion are denoted by a straight line. (c) Photoluminescence emission spectra of the sample S120 with different excitation wavelengths. (d) Photoluminescence spectra of all the samples with excitation wavelength 330 nm.

*3.2.1. Photoluminescence Emission Spectra Analyses*

The photoluminescence (PL) peak represents the radioactive transition between two energy states. Here, at first, the room temperature photoluminescence spectra of S120 sample are taken with different excitations at 300 nm, 330 nm, 350 nm, 370 nm and 400 nm and the spectra are represented logarithmically in **Figure** 5 **(c)**. Primarily, the results indicate the poor luminescent feature of synthesized S120 sample. However, in all the cases, an emission peak at ~468 nm is observed. This peak at ~468 nm can be assigned as the band-to-band transition [31, 40, 61]. Further, PL spectra of all the synthesized samples are taken with 330 nm excitation and represented in **Figure** 5 **(d)**. Here again, only a prominent signal at ~468 nm with varying intensities is noticed along with a weak PL signal at ~ 545 nm. As said earlier, the presence of the most intense peak at ~468 nm for all the cases is due to the band to band transition i.e. transition between W5d orbital (CB) and hybrid orbital of Sb6s and O2p (VB) in the $WO_6$ [62]. The presence of small emission at ~545 nm is due to defects related sub-band gap level transition. Note that, absorption study of the ATN possesses different band tailing indicating presence of either various types of defects or defects with various concentration. However, it is not reflected in the PL spectra indicating the involved defects sites in $Sb_2WO_6$.

## 4. Photocatalytic Dye Degradation Study

Photocatalytic degradation of MB dye of ATN using samples S120, S140, S160, S180 and S200 as catalysts are done in the aqueous medium of pH 6.7 during 180 min visible light (LED) irradiation. The variation of the UV-Vis absorption spectra of the MB dye remaining in the solution at different instant of time are shown in **Figure S1** (supplementary data). It is clear from the plot that the intensity decreases consistently with time. For MB dye, the concentration is taken to be proportional to the intensity at the peak, 663 nm. **Figure 6(a)** represents $C/C_0$ versus irradiation time (t) plot for degradation of MB in the presence of the catalysts. Where $C_0$ is the initial dye concentration and C is the concentration of the dye remaining non-degraded after a certain time of irradiation. The degradation efficiency is calculated using the formula

$$Degradation\ (\%) = \left(1 - \frac{C}{C_0}\right) \times 100\%$$

The sample S180 shows the highest degradation efficiency, almost 98% of the dye degraded within 120 min of visible irradiation. While it is well documented that in absence of catalyst under visible light irradiation almost 97% dye remains non-degraded up to 180 min [9]. The $\log(C_0/C)$ versus time (t) plot of these materials is represented in **Figure 6(b)**. Moreover, rate constant, considering pseudo first-order kinetics, is determined using the equation $\log(C_0/C)=kt$, where k is the degradation rate constant. The degradation rate constant calculated from the linear fitting of the logarithmic plot in presence of the samples S120, S140, S160, S180 and S200 are $0.00137\pm0.0001$ min$^{-1}$, $0.00349\pm0.00015$ min$^{-1}$, $0.00492\pm0.00025$ min$^{-1}$,

0.015±0.00109 min$^{-1}$, 0.01148±0.0007 min$^{-1}$ respectively. For comparision, k values of other samples with Sb$_2$WO$_6$ as base materials are shown in Table IV.

From the above results it is evident that the photocatalytic efficiency consistently increases from S120 with the highest efficiency is observed for the sample S180, while the efficiency decreased slightly for S200. Here, the factors behind this variation is the change in visible light absorption, which has been shown in Figure 4(a). The other factors are the distortion of the atoms in the crystals giving rise to internal electric field affecting the electron-hole recombination and bandgap variation.

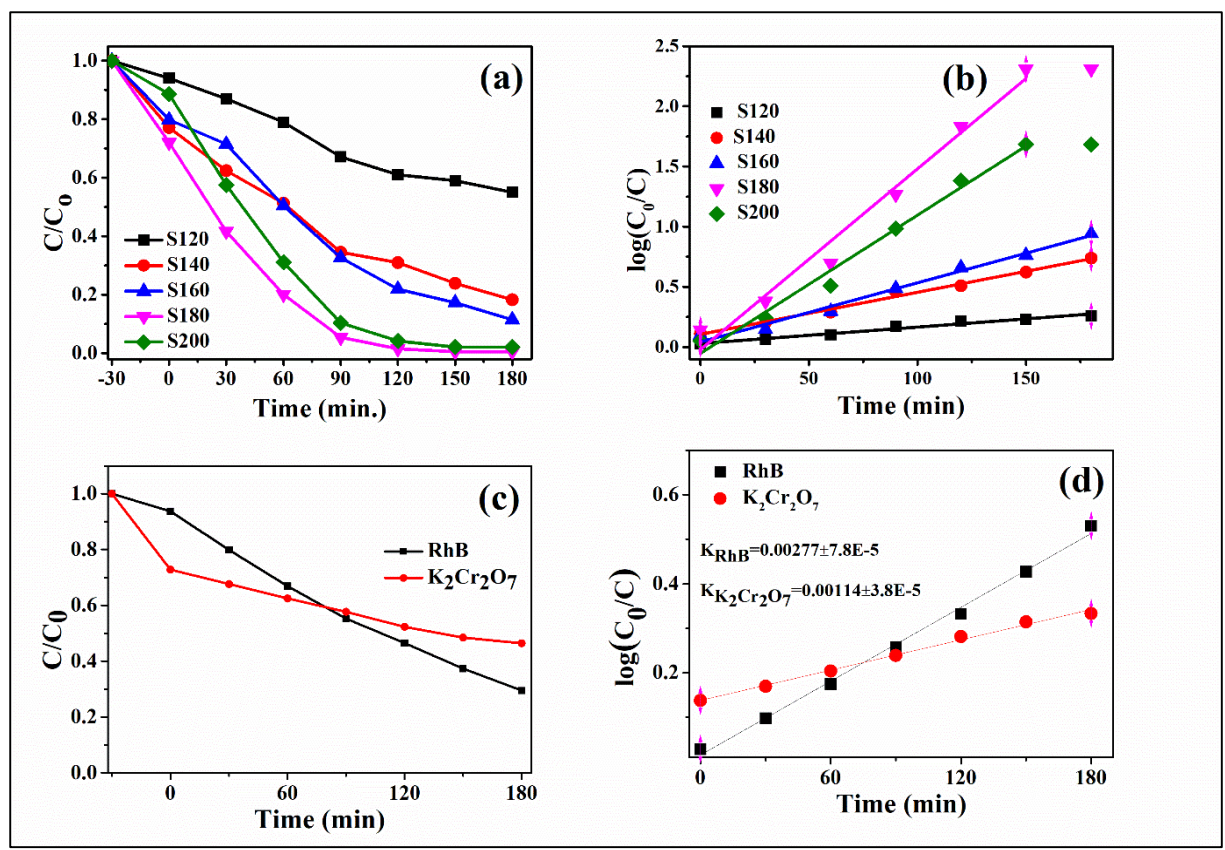

**Figure 6: (a)** C/C$_0$ versus time plot for MB degradation using the samples S120, S140, S160, S180 and S200 . (b) log(C$_0$/C) versus time plot with straight line fitting. (c) Photocatalytic efficiency of the S180 for degrading RhB and K$_2$Cr$_2$O$_7$. (d) Straight line fitted curve of log(C$_0$/C) versus time.

Having the highest photocatalytic efficiency, the sample S180 is further used for the degradation of RhB and K$_2$Cr$_2$O$_7$ under same conditions. Both of these are stable compound, degradation efficiency of the S180 sample is shown in the **Figure 6(c)**. Within 180 min of visible light irradiation 5 mg S180 sample can degrade 70% of the RhB dye and 53% of $K_2Cr_2O_7$. Considering photocatalytic effect as first order reaction the rate constants are 0.00277 min$^{-1}$ and 0.00114 min$^{-1}$ respectively.

**Table IV**: Photocatalytic activity of Sb2WO6 reported earlier, where different dyes, for example RhB, MB and Methyl Orange (MO) are used as pollutants.

| Sample | Dye used | Photocatalytic efficiency | Reference |
| --- | --- | --- | --- |
| 3DGA/Sb$_2$WO$_6$ | MO | 57.8% in 5 hours | [51] |
| RGO-Sb$_2$WO$_6$ | RhB | 91% in 80 min | [36] |
| GQD/Sb$_2$WO$_6$ | MO | 57.8% in 5 hours | [63] |
| Ag$_2$WO$_4$/Sb$_2$WO$_6$ | RhB | 80.8% in 90 min | [33] |
| 1%CQD/Sb$_2$WO$_6$ | RhB | 83% in 120 min | [64] |
| Sb$_2$WO$_6$ nano particle | MB | 98% in 120 min | This work |

Other than the efficiency, applicability towards its industrial as well as scientific purpose of any sample depends on its repeated use. Hence the stability, more specifically photostability, and reusability of the sample is a crucial factor for any sample to be physically applicable. Photostability of the ATN is inspected by degrading MB under identical conditions over consecutive 5 cycles. The photocatalytic efficiency of S180 in each cycle is represented by bar diagram in **Figure 7(a)**. It is observed that the sample S180 shows appreciable resistant towards photocorrosion, retaining over 85% efficiency at the end of 5$^{th}$ cycle. The similarity in XRD pattern before and after 5 consecutive photocatalytic cycles shows that there is no deformation in the crystal structure in presence of light. Though the peak positions of the XRD pattern before and after cyclic test remain same, as shown in **Figure 7(b,c)**, an increase in background noise in observed. The background noise arises due to the residual chemicals after MB degradation and also the surface deformation during photocatalysis.

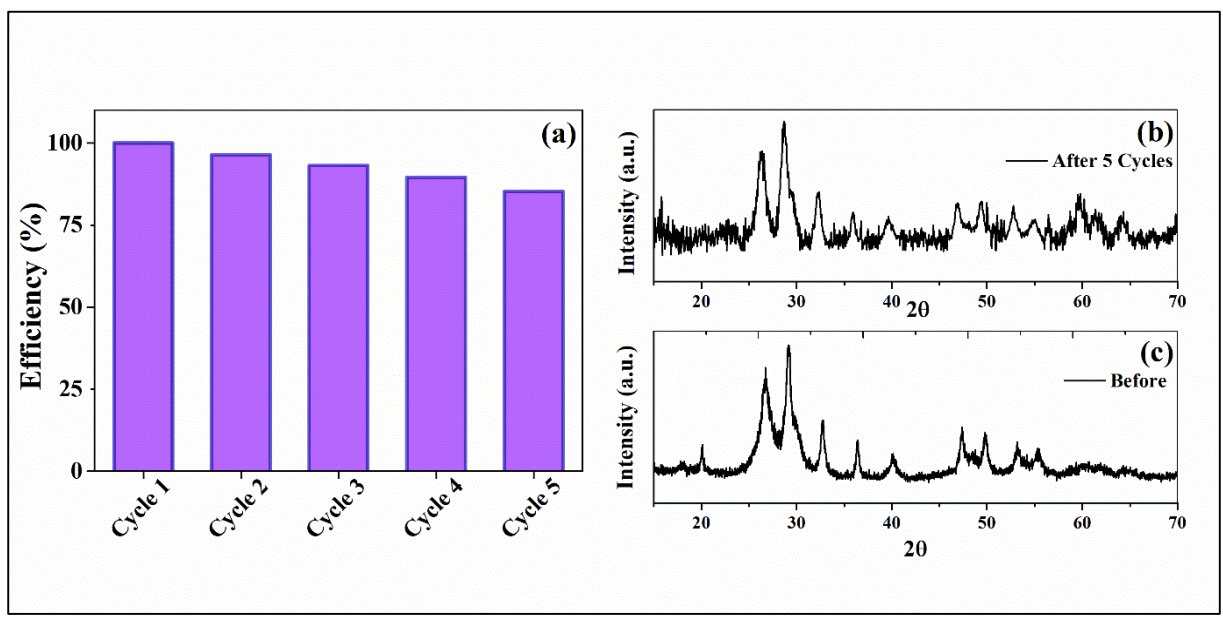

*Figure 7: (a) Efficiency of the sample S180 after each cycle. (b) XRD pattern after 5$^{th}$ cycle (c) Initial XRD pattern.*

## 5. Conclusion:

In summary, a batch of ATN is synthesized through a simple, cost-effective one-step hydrothermal process by varying the synthesis temperature in the range of the 120 °C to 200 °C. XRD data reveal their polycrystalline nature and the crystal sizes are found to be increased with synthesis temperature. Moreover, W-H plots of the said samples reveal the presence of compressive strain inside the crystals, while Rietveld analysis gives the information about the temperature induced cell volume variation of the crystal. Further, some new Raman active modes of the $Sb_2WO_6$ are identified along with a consistent change in few Raman modes as a result of temperature variation during synthesis. Again the series of samples show a significant variation in the optical properties in terms of bandgap calculated from UV-Vis spectra and luminescent properties. In particular, sample S180 is found to possess the lowest value equal to 2.38 eV having a significant absorptivity in the visible region. Photoluminescence spectra of all the synthesized samples reveal weak luminescence, however gives an emission around ~468 nm that corresponds to band to band transition. Therefore, this work has demonstrated clearly that an effective bandgap tuning of $Sb_2WO_6$ is possible by varying the synthesis temperature in hydrothermal process based synthesis. The formation of the crystal defects that are associated with sub bandgap level formation is responsible for the effective bandgap decrease of the material. Besides, ATN is prepared by easiest one step hydrothermal process showcase remarkable photocatalytic effect by degrading MB in 180 minutes almost completely while RhB has been degraded about 70% in 180 minutes under visible light (LED) irradiation. The sample S180 is also used for degrading $K_2Cr_2O_7$ showing an efficiency of approximately 53% within 180 min and a corresponding rate constant of 0.00114 min$^{-1}$. We believe that this work and the related observations will help researchers for exploring the visible light photocatalytic property of heterostructured and functionalized antimony tungstate.

Acknowledgements:

The authors, D. Karmakar and S. K. Mandal want to thank Council of Scientific and Industrial Research (CSIR), Government of India and S. Paul acknowledge SERB (National Post Doctoral Fellowship), India for providing fellowship during the tenure of this work.